\documentclass[10pt]{iopart}

\usepackage{amssymb}
\usepackage{graphicx}
\usepackage{cite}
\usepackage{array}

\begin{document}

\title{Transport, shot noise, and topology in AC-driven  dimer arrays}

\author{Michael Niklas$^1$, M\'onica Benito$^2$, Sigmund Kohler$^2$ and Gloria Platero$^2$}

\address{$^1$ Institut f\"ur Theoretische Physik, Universit\"at Regensburg,
93040 Regensburg, Germany}
\address{$^2$ Instituto de Ciencia de Materiales de Madrid, CSIC, 28049
Madrid, Spain}

\begin{abstract}
We analyze an AC-driven dimer chain connected to a strongly biased electron
source and drain.  It turns out that the resulting transport exhibits
fingerprints of topology.  They are particularly visible in the
driving-induced current suppression and the Fano factor.  Thus, shot noise
measurements provide a topological phase diagram as a function of the
driving parameters.  The observed phenomena can be explained physically by
a mapping to an effective time-independent Hamiltonian and the emergence of
edge states.  Moreover, by considering quantum dissipation, we determine
the requirements for the coherence properties in a possible experimental
realization.  For the computation of the zero-frequency noise, we develop
an efficient method based on matrix-continued fractions.
\\[3ex]
\today
\end{abstract}

\pacs{05.60.Gg,         
      03.65.Vf,         
      73.23.Hk          
}
\vspace{2pc}
\noindent{\it Keywords}: Quantum transport, time-dependent quantum systems, topology

\maketitle
\ioptwocol

\section{Introduction}

The ever smaller size of quantum dots implies small capacitances and
accordingly large charging energies.  Indeed in most recent realizations of
coupled quantum dots, Coulomb repulsion represents the largest energy scale
\cite{vanderWiel2003a, Taubert2011a} such that states with different
electron numbers are energetically well separated.  Then the quantum dot
array can be controlled by gate voltages and, despite a possible
coupling to electron reservoirs, the dynamics is restricted to a few states
with a specific electron number.  This forms the basis for many
realizations of spin or charge qubits.

While quantum information processing is usually performed in closed
systems, the possibility to couple quantum dots to electron source and
drain may be useful as well.  It not only can be exploited for qubit
readout \cite{Elzerman2004a}, but also allows one to determine the relevant
system parameters.  For example, upon increasing the source-drain bias, an
increasing number of levels enters the voltage window such that a current
measurement provides the spectrum of the coupled quantum dots.  The
dominating feature in the current-voltage profile is provided by the
mentioned charging effects which cause Coulomb blockade.  Further blockade
mechanisms come about when in addition spin effects \cite{Weinmann1995a} or
phononic excitations \cite{Weig2004a, Koch2005a, Hubner2009a} play a role.
Moreover, in a dimer chain, the interplay of Coulomb interaction and the
topology may cause edge-state blockade \cite{Benito2016a}.  It is
particularly visible in the shot noise properties of the transport process.

\begin{figure}
\centerline{\includegraphics[width=\columnwidth]{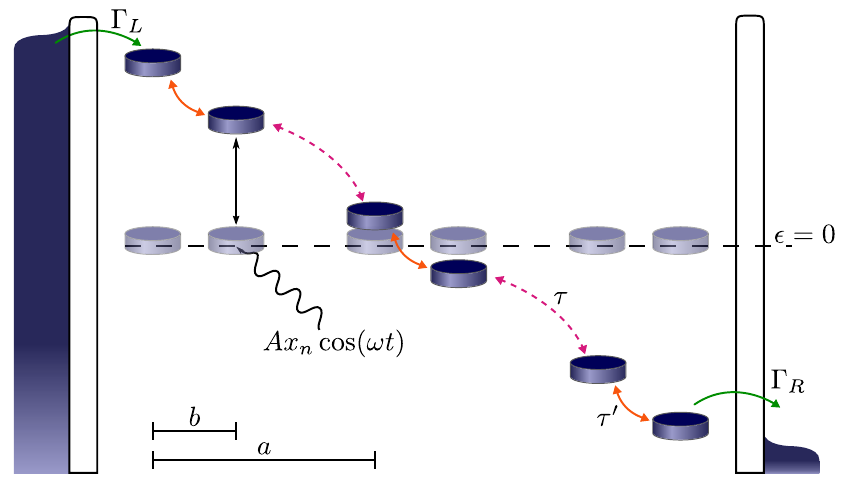}}
\caption{Dimer chain with intra-dimer tunnel coupling
$\tau'=\tau_0-\delta\tau$ and inter dimer coupling
$\tau=\tau_0+\delta\tau$, connected to the electron source (left) and drain
(right) with rate $\Gamma_{L,R}$. The applied external AC field generates
an oscillation of the onsite energies with frequency $\omega$ which depends
on the positions of the quantum dots, $x_n$, therefore the inter and intra dimer
distances, $a$ and $b$ respectively, become relevant even for the
tight-binding description.
}
\label{fig:setup}
\end{figure}

Experimental evidence for edge-state blockade will be facilitated by a high
tunability of the inter-dot tunneling.  A possible way to circumvent this
problem is driving the conductor by an electric dipole field.  Then for not
too small frequencies, the driving essentially renormalizes the inter-dot
tunnel coupling \cite{Grossmann1991a, Grossmann1992a, Holthaus1992b,
Creffield2002a} and, thus, allows the emulation of a dimer chain with
highly tunable tunneling.  In the corresponding transport setting, i.e., in
the presence of electron source and drain, one expects a corresponding
current suppression \cite{Platero2004a, Kohler2005a} which
has been measured in double quantum dots \cite{Stehlik2012a, Forster2014a}.
Moreover, the driving may have significant impact on the shot noise
\cite{Camalet2003a}.
In this work, we explore the possibility for edge-state blockade in ac
driven quantum dot arrays such as those sketched in Fig.~\ref{fig:setup}.
It is based on the recent finding that the topological properties of ac
driven dimer chains can be controlled via the amplitude of a driving field
\cite{GomezLeon2013a, DalLago2015a, Bello2016a}.

In the regime of strong Coulomb repulsion and relatively small dot-lead
tunneling, an established way to describe transport are master equations of
the Bloch-Redfield type \cite{Redfield1957a, Blum1996a, Breuer2003a}.  In
combination with a Floquet theory for the central system, they can be
applied to periodically driven transport problems including to the
computation of shot noise \cite{Kohler2005a}.  The computation of the
required dissipative kernels, however, may be numerically demanding.  In
certain limits, however, the master equation assumes a convenient Lindblad
form which is easier to evaluate.  For its efficient numerical treatment,
we extend a previously developed matrix-continued fraction method
\cite{Forster2015b} to the computation of shot noise.

This work is structured as follows.  In Sec.~\ref{sec:model}, we introduce
our model and a master equation description, as well as a
matrix-continued fraction method for the computation of the current and the
zero-frequency noise of time-dependent transport problems. The main
features of the current are presented in Sec.~\ref{sec:transport}, while
Sec.~\ref{sec:dissipation} is devoted to the impact of dissipation.  In the
\ref{sec:I(t)}, we test an assumption made for the efficient computation of
the Fano factor.

\section{Model and master equation}
\label{sec:model}

We consider an array of quantum dots coupled to a time-dependent electric
dipole field such that the onsite energies oscillate in time with a
position-dependent amplitude, see Fig.~\ref{fig:setup}.  The static part is
given by the Su-Schrieffer-Heeger (SSH) Hamiltonian \cite{Su1979a}
\begin{equation}
\label{ssh}
H_\mathrm{SSH} = \sum_{n=1}^{N-1} \tau_n c_{n+1}^{\dagger} c_{n} + \mathrm{h.c.} ,
\end{equation}
where $c_n$ is the fermionic annihilation operator for an electron on site
$n$ located at position $x_n$.  We will focus on a dimer chain with an even
number of sites, $N$, and alternating tunnel matrix elements
$\tau_n = \tau_0+(-1)^n\delta\tau$.  The AC field couples via the dipole
operators of the chain such that the Hamiltonian of the driven chain reads
\begin{equation}
\label{eq:model}
H(t) = H_\mathrm{SSH} + A\sum_{n=1}^{N} x_n c_{n}^{\dagger} c_n \cos(\omega t)
\end{equation}
with $x_n$ the position of site $n$.  The distances between two neighboring
sites are $b$ and $a-b$, which implies a unit cell of length $a$.
The driving is determined by its frequency $\omega$ and amplitude $A$.

\subsection{Generalized chirality}

The topological properties of the SSH model stem from a chiral (or
sub-lattice) symmetry $C$ with $CH_\mathrm{SSH}C = -H_\mathrm{SSH}$.  In
second quantization, this symmetry operation can be written as
\begin{equation}
C = e^{\pi i\sum_n n c_n^\dagger c_n} = C^{-1} .
\end{equation}
In essence, it provides a minus sign for all creation and annihilation
operators with an odd site number since
\begin{equation}
C c_n C = (-1)^n c_n .
\end{equation}
Consequently, all nearest-neighbor hoppings acquire a factor $-1$ which
explains the mentioned chiral symmetry of $H_\mathrm{SSH}$.  The
time-dependent part of the Hamiltonian \eref{eq:model} consists of local
terms $c_n^\dagger c_n$ which are invariant under transformation with $C$.
However, the sinusoidal driving allows us to obtain a minus sign via
shifting the time by half a driving period, $t\to t+T/2$, where
$T=2\pi/\omega$.  Formally, this can be expressed as
\begin{equation}
\label{eq:GC}
C H(t) C = -H(t+T/2) .
\end{equation}
We refer to this symmetry relation as ``generalized chirality'', owing to
its resemblance to the generalized parity present in symmetric bistable
potentials driven by a dipole force \cite{Grossmann1991a}.

A consequence of the generalized chirality is that the propagator of the
chain, $U(t,t')$, obeys the relation
\begin{equation}
U(t+T,t+T/2) = C U^{-1}(t+T/2,t) C .
\end{equation}
Thus, the one-period propagator can be split in two symmetry-related parts,
a fact that has been identified as a condition for non-trivial topological
properties of a periodically driven system \cite{Asboth2014a}.

\subsection{Counting statistics}

Transport is enabled by coupling the first and the last site to an electron
source and drain, respectively.  In the limit in which the applied voltage
is much larger than the tunnel matrix elements $\tau_n$, but still
considerably smaller than the Coulomb repulsion of the electrons on the
array \cite{Benito2016a}, a standard Bloch-Redfield approach to second
order in the chain-lead tunneling provides the Lindblad master equation
\begin{equation}
\label{eq:meq}
\dot\rho = \mathcal{L}\rho = -\frac{i}{\hbar}[H(t),\rho]
+ \Gamma_L \mathcal{D}(c_1^\dagger)\rho +\Gamma_R \mathcal{D}(c_N)\rho ,
\end{equation}
with $\mathcal{D}(x)\rho = (2x \rho x^\dagger - x^\dagger x\rho - \rho
x^\dagger x)/2$ and the dot-lead rates $\Gamma_{L,R}$.  In the limit
considered, Eq.~\eref{eq:meq} has to be evaluated in the basis with at most
one electron on the array.  This restricted basis captures Coulomb
interaction.

To compute the statistics of the transported electrons, we have to
generalize the master equation by introducing a counting variable $\chi$
for the electrons in the right lead.  Proceeding as in
Ref.~\cite{Benito2016a}, we consider the generalized reduced density
operator $R_{\chi}$ which for the present case of uni-directional transport
obeys the equation of motion \cite{Bagrets2003a}
\begin{equation}
\dot{R}_{\chi}(t) = \left[\mathcal{L} +(e^{i\chi}-1)\mathcal{J}\right] R_{\chi}(t)\,.
\end{equation}
It is defined such that the moment-generating function of the electron
number in the right lead, $N$, becomes $ \langle e^{i\chi N}\rangle =\tr
R_\chi$ and, thus, $\langle N^k\rangle_t = (\partial/i\partial\chi)^k \tr
R_\chi(t)|_{\chi=0}$.  Taylor expansion of $R_\chi$ in $\chi$ as
\begin{equation}
R_{\chi}(t) = F_0(t) + \sum_{k=1}^{\infty}\frac{(i\chi)^k}{k!}F_k (t)
\end{equation}
relates the $F_k$ to the moments $\langle N^k\rangle_t = \tr F_k(t)$.
Inserting this decomposition into the generalized master
equation \eref{eq:meq}, provides the hierarchy
\begin{eqnarray}
\dot F_0 &=& \mathcal{L} F_0 \label{eq:F0} \,,\\
\dot F_1 &=& \mathcal{L} F_1+ \mathcal{J} F_{0}\label{eq:F1} \,, \\
\dot F_2 &=& \mathcal{L} F_2+\mathcal{J} F_{0}+ 2\mathcal{J} F_{1}\label{eq:F2} \,.
\end{eqnarray}
Obviously, the first equation is the original master equation \eref{eq:meq}
and $F_0=\rho$.

For typical Markovian and time-independent transport problems, the
cumulants eventually grow linearly in time, which motivates us to focus on
the time-derivatives $\tr\dot F_k = (d/dt)\langle N^k\rangle$.  To
obtain the current expectation value, we solve Eq.~\eref{eq:F0} and insert
the solution into Eq.~\eref{eq:F1} which yields
$I(t) =\tr\dot{F}_{1}=\tr \mathcal{J}F_{0}$.
The second moment follows from Eq.~\eref{eq:F2} together with the solution
of the first two equations.  Subtracting the time-derivative of $\langle
N\rangle^2$, we obtain the zero-frequency noise
\begin{equation}
S(t)=\frac{d}{dt}\left[\tr F_2-(\tr F_1)^2\right]=I(t)+2\tr \mathcal{J} F_{1\perp} \,,
\end{equation}
where $F_{1\perp}=F_1-F_0\tr F_1$ represents the component of $F_1$
perpendicular to $F_0$.  It obeys the equation of motion
\begin{equation}
\dot{\mathcal{F}}_{1\perp} = \mathcal{L}\mathcal{F}_{1\perp}
+\left(\mathcal{J}-I(t)\right) F_{0}\ .\label{eq:F1perp}
\end{equation}
Notice that this equation depends on the current expectation value $I(t)$
and, thus, on the solution of the master equation \eref{eq:meq}.

\subsection{Matrix-continued fractions}
\label{sec:mcf}

A natural way to solve Eqs.~\eref{eq:meq} and \eref{eq:F1perp} is the
numerical integration of the first equation followed by the computation of
$I(t)$ and the numerical integration of the second equation.  While being
very flexible, such numerical propagation schemes often lack
efficiency.  Therefore we aim at implementing a matrix-continued fraction
method \cite{Risken1989a} which in the context of mesoscopic transport has
been employed recently for the computation of time-averaged currents
\cite{Forster2015b}.  Here we extend this scheme to the computation of the
zero-frequency noise.

For convenience, we write the two equations of motion in block matrix
notation,
\begin{eqnarray}
\left(\begin{array}{c} \dot{\rho}\\ \dot{{\cal F}}_{1\perp}
\end{array}\right)
&=& \left(\begin{array}{cc}{\cal L}(t) & 0\\
     {\cal J}-I(t) & {\cal L}(t)
  \end{array}\right)
  \left(\begin{array}{c}\rho \\ {\cal F}_{1\perp}
  \end{array}\right)
\\ &\equiv&  M(t) \mathbf{b}
\label{eq:eqM}
\end{eqnarray}
with the shorthand notation $\mathbf{b}=(\rho,\mathcal{F}_{1\perp})^{T}$.
To derive a matrix-continued fraction scheme, we have to bring this
equation into the form of a tridiagonal recurrence relation
\cite{Risken1989a}.  In the present case, this is hindered by the fact that
$M(t)$ depends on the time-dependent current $I(t)$ which may contain
higher-order harmonics.  Here however, we find that reliable results for
the noise can still be obtained when $I(t)$ is replaced by its time
average.  We test this assumption in \ref{sec:I(t)}.
Since now the remaining time-dependence in $M(t)$ stems from the Liouvillian of
the sinusoidal driving in the Hamiltonian \eref{eq:model}, the Fourier
decomposition of the terms in Eq.~\eref{eq:eqM} reads
\begin{eqnarray}
M(t) &=& M_0+M_+ e^{i\omega t}+M_- e^{-i\omega t}\label{eq:Mt} , \\
\mathbf{b}(t)&=&\sum_{n=-\infty}^{\infty}e^{i n \omega t}\mathbf{b}_n .
\label{eq:bt} 
\end{eqnarray}
By inserting Eq.~\eref{eq:bt} into Eq.~\eref{eq:eqM} we obtain the tridiagonal
recurrence relation
\begin{equation}
M_+\mathbf{b}_{n-1}+\left(M_0-in \omega \right)\mathbf{b}_n+M_-\mathbf{b}_{n+1}=0\,. \label{eq:recurrence}
\end{equation}

Our interest lies in the time-average of $\mathbf{b}(t)$, i.e., in the
Fourier component $\mathbf{b}_0$.  To this end, we define the transfer matrices
$S_k$ and $R_k$ via the ansatz
\begin{equation}
\label{defSR}
\mathbf{b}_n =
\cases{
R_n  \mathbf{b}_{n+1} & for $n<0$ ,
\\
S_n  \mathbf{b}_{n-1} & for $n>0$ .
}
\end{equation}
Consistency with Eq.~\eref{eq:recurrence} is ensured by the recurrence
relations
\begin{eqnarray}
\label{eq:Sn}
S_n ={}& -\left[M_0 -i n\omega +M_- S_{n+1}\right]^{-1}M_+ ,
\\
\label{eq:Rn}
R_n ={}&  -\left[M_0 -i n\omega +M_+ R_{n-1}\right]^{-1}M_- ,
\end{eqnarray}
together with
\begin{eqnarray}
\label{app:p0}
(M_0+M_+ R_{-1} + M_- S_1)\mathbf{b}_0 = 0 . \label{eq:b0}
\end{eqnarray}
For practical purposes, we have to truncate the Fourier components of
$\mathbf{b}(t)$ assuming $\mathbf{b}_n=0$ for $|n|>n_0$ which holds for
$S_{n_0+1}=R_{-(n_0+1)}=0$.  With the latter condition we compute $R_{-1}$
and $S_1$ by iterating Eqs.~\eref{eq:Sn} and \eref{eq:Rn} which finally
provides an explicit expression for Eq.~\eref{eq:b0}.  In a last step we
solve this homogeneous equation under the trace conditions $\tr\rho_0=1$
and $\tr\mathcal{F}_{1\perp}=0$.

\section{Transport in the high-frequency regime}
\label{sec:transport}

The main energy scale of the SSH Hamiltonian \eref{ssh} is the bandwidth
$\tau_0$.  If it is much smaller than the energy quanta of the driving
field, $\hbar\omega$, one may employ a high-frequency approximation to
derive an effective time-independent Hamiltonian that captures the
long-time-dynamics of the driven system.  This typically results in an
effective Hamiltonian with parameters renormalized by Bessel functions.  In
this way, the AC-driving offers a possibility for tuning system parameters.
A classic example is the suppression of tunneling in bistable potentials
\cite{Grossmann1991a, Grossmann1992a} and superlattices
\cite{Holthaus1992b, Platero1997a} by the purely coherent influence of an
ac field.

In a dimer chain driven by an external electric field, the  intra and
inter dimer spacings become relevant because they determine the dipole
moments and, thus, appear in the renormalizations of the inter dimer
tunneling $\tau$ and the intra-dimer tunneling $\tau'$ which in our case
read
\begin{eqnarray}
\label{eq:tau'}
\tau_\mathrm{eff}' &=& J_0(Ab/\omega) (\tau_0-\delta\tau)  , \\
\label{eq:tau}
\tau_\mathrm{eff} &=& J_0(A(a-b)/\omega) (\tau_0+\delta\tau) ,
\end{eqnarray}
where $J_0$ is the zeroth-order Bessel function of the first kind.  For
details of the calculation, see Ref.~\cite{GomezLeon2013a}.  With these
effective tunnel matrix elements, one can draw conclusions about the
topological properties of the chain by a comparison with results for the
time-independent SSH model \cite{Zak1989a, Delplace2011a}. The main finding
is a trivial topology for $\tau_\mathrm{eff}'> \tau_\mathrm{eff}$, while
for $\tau_\mathrm{eff}'< \tau_\mathrm{eff}$ it becomes non-trivial with a
Zak phase $\pi$ \cite{GomezLeon2013a, DalLago2015a}.  Similar influence of
radiation on topology occurs also in higher dimensions
\cite{Lindner2011a, Grushin2014a, Usaj2014a}.

The tunnel matrix elements \eref{eq:tau'} and \eref{eq:tau} possess an
interesting duality.  By the replacement $(\delta\tau, b) \to
(-\delta\tau,a-b)$, these matrix elements are interchanged.  Then the
topological properties are interchanged as well, while the bulk spectra
remain the same.  This motivated the choice of parameters used in
Fig.~\ref{fig:cdt}.

\subsection{Current suppression and edge-state blockade}

\begin{figure}
\centerline{\includegraphics[width=\columnwidth]{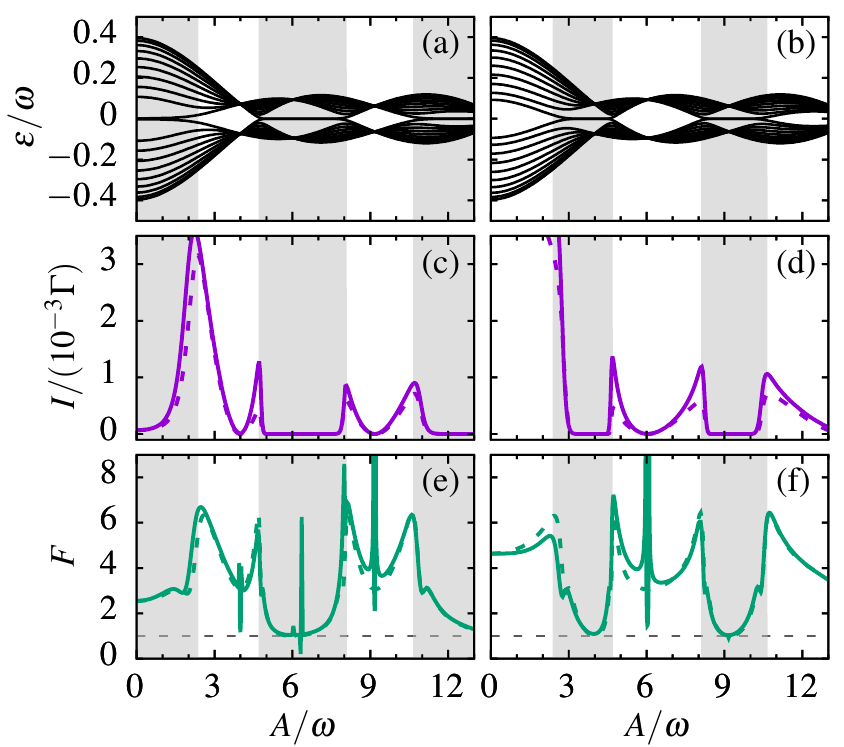}}
\caption{(a,b) Quasienergy spectrum as a function of the driving amplitude
$A$ for (a) $\delta\tau=0.2\tau_0$, $b=0.4a$ and (b) $\delta\tau=-0.2\tau_0$,
$b=0.6a$ for a chain with $N=20$ sites.  The parameters are chosen such
that the bulk spectra in the thermodynamic limit are identical, while the
topological properties depend on the sign of $\delta\tau$.
Accordingly, we find edge-states at zero quasienergies in the regions marked by
a grey background.  (c,d) Time-averaged current.  The dashed lines
correspond to the high-frequency approximation result.  The driving
frequency is $\omega=5\tau_0$, while the chain-lead coupling reads
$\Gamma=5\tau_0$.  (e,f) Fano factor $F=\bar{S}/\bar{I}$.}
\label{fig:cdt}
\end{figure}

If the driving amplitude $A$ is such that one of the Bessel functions in
Eqs.~\eref{eq:tau'} and \eref{eq:tau} vanishes, the way from the electron
source to the drain is practically interrupted, which significantly reduces
the current.  The data in Fig.~\ref{fig:cdt} confirm this
expectation and reveal a particular dependence on topology: It is best
visible in a comparison of data for two parameter sets that are related by
the transposition of inter-dot and intra-dot coupling (left and right
column, respectively, of this figure).  Both choices lead to the same bulk
properties, while the topological and the trivial regions are interchanged.
This allows us to identify topological effects.  The complementarity of
the two cases is evident from the quasienergy spectra shown in
Figs.~\ref{fig:cdt}(a) and \ref{fig:cdt}(b).

Figures~\ref{fig:cdt}(c) and \ref{fig:cdt}(d) show a remarkable dependence
of the current suppression on the topology. In the trivial region, the
current is extensively reduced only when the effective inter dimer
tunneling vanishes, i.e., for $\tau_\mathrm{eff}\ll\tau_0$ [$A\approx
9\omega$ in Fig.~\ref{fig:cdt}(c) and $A\approx 6\omega$ in
Fig.~\ref{fig:cdt}(d)].  Close to the suppression, the current grows
quadratically, such as for a driven double quantum dot \cite{Kohler2005a}.
By contrast, the current almost vanishes in the whole topological region,
i.e., whenever the weaker condition
$\tau_\mathrm{eff}'<\tau_\mathrm{eff}$ is fulfilled.  Therefore, we can
conclude that the physical origin of this current suppression is not a
completely vanishing effective tunnel matrix element, but must be related
to topology and the corresponding edge states formed at the source and at
the drain.  These edge states possess two characteristic features. First,
they are exponentially weakly connected and, second, they
are energetically well separated from the bulk states.  As a consequence,
they may trap electrons and thereby interrupt the transport process such
that one observes \textit{edge-state blockade}.  As compared to its
counterpart in time-independent chains \cite{Benito2016a}, this blockade is
characterized by a broad region with vanishing current, while the CDT-like
suppression of current in the trivial region has a parabolic shape.

\subsection{Shot noise and phase diagram}

For less tunable static chains, it has been proposed to identify edge-state
blockade by its characteristic shot noise properties \cite{Benito2016a}.
In particular, it has been found that the small current in the blockade
regime obeys Poissonian statistics ($F\approx 1$), while the transport in the
trivial regime is characterised by electron bunching \cite{Blanter2000a,
Emary2012a}.  Figures~\ref{fig:cdt}(e) and
\ref{fig:cdt}(f) depict the shot noise for the driven case characterized by
the Fano factor.  It reveals a smeared crossover between Poissonian noise
and super Poissonian values up to $F\approx 8$.

\begin{figure}
\centerline{\includegraphics[width=\columnwidth]{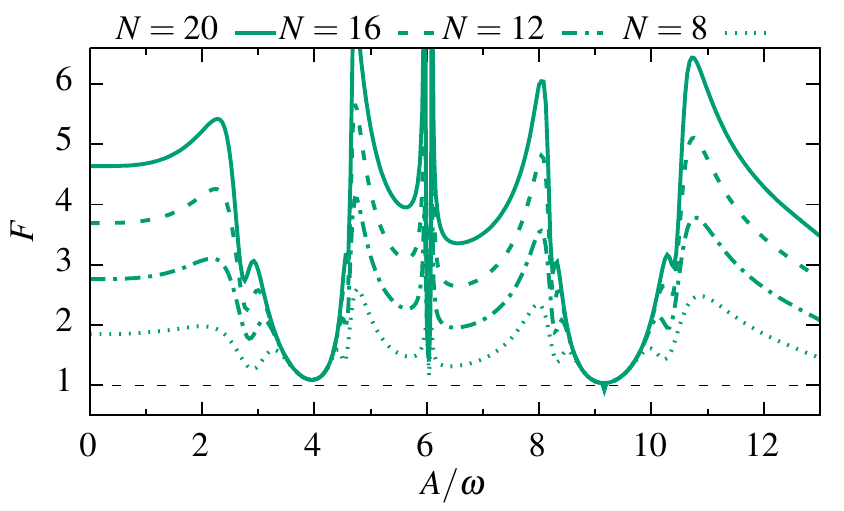}}
\caption{Fano factor $F=\bar{S}/\bar{I}$ as a function of the driving
amplitude $A$ for $\delta\tau=-0.2\tau_0$, $b=0.6a$ for chains of
various lengths.  The driving frequency and the lead-chain coupling are
$\omega=5\tau_0$ and $\Gamma=5\tau_0$, respectively.}
\label{fig:Ndependence}
\end{figure}

The difficulty of performing an experiment on a chain with many sites
raises the question about the necessary length to observe the \textit{edge
state blockade}. Thus we have calculated the Fano factor corresponding to
the parameters in Fig.~\ref{fig:cdt}(f) for chains of different length. An
advantage of using the external AC field to manipulate the topological
phase is that the ideal Poissonian Fano factor $F\approx1$ is always
reached for a certain point in the blockade region, as shown in
Fig.~\ref{fig:Ndependence}.  This finding is in contrast to the static
case, where $F\approx1$ was found only in the limit of very long chains
\cite{Benito2016a}. However, Fig.~\ref{fig:Ndependence} also shows that for
a short chain the Fano factor in the CDT point also approaches unity, which
does not allow distinguishing the CDT effect from the topological blockade.

\begin{figure}
\centerline{\includegraphics[width=\columnwidth]{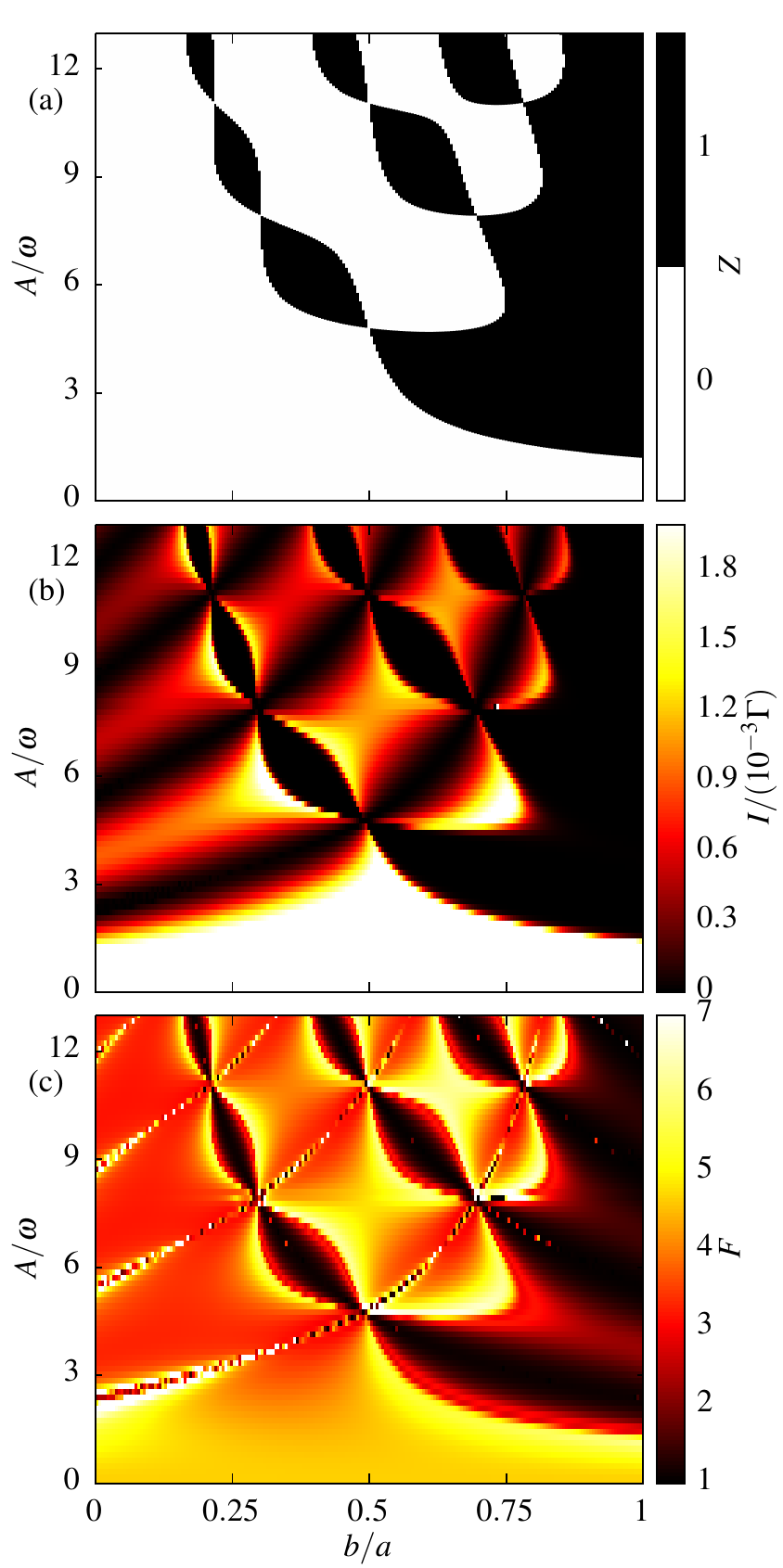}}
\caption{(a) Chern number as calculated in Ref.~\cite{GomezLeon2013a} as a
function of the intra dimer distance $b$ and the driving amplitude $A$.
All other parameters are as in the right column of Fig.~\ref{fig:cdt}.
(b) Corresponding time-averaged current and (c) Fano factor.}
\label{fig:phasediagram}
\end{figure}

While in contrast to the static case \cite{Benito2016a}, here the shape of
the current suppression may be sufficient to identify edge-state blockade,
it will turn out that the Fano factor exhibits clearer fingerprints of the
topological phase diagram computed as in Ref.~\cite{GomezLeon2013a} and shown
in Fig.~\ref{fig:phasediagram}(a).  The corresponding plot for the current
[Fig.~\ref{fig:phasediagram}(b)] exhibits a richer structure stemming from
the additional current suppressions in the trivial regions.  Therefore the
behavior of the current alone does not reflect the topological phase.  The
Fano factor [Fig.~\ref{fig:phasediagram}(c)], by contrast, provides clearer
evidence, because $F\approx1$ is found exclusively for non-trivial topology
(black regions).  We also find some additional structure in the trivial
region as narrow lines at CDT-like zeros of the current.  There, the Fano
factor assumes even larger values which correspond to the sharp peaks in
Figs.~\ref{fig:cdt}(e) and \ref{fig:cdt}(f).  Thus, shot noise measurements
represent an alternative to the direct observation of the Zak phase
\cite{Atala2013a}.

\section{Quantum dissipation}
\label{sec:dissipation}

In Ref.~\cite{Benito2016a}, we have shown that for the static SSH model,
the fingerprints of the topological properties in the Fano factor are
fairly insensitive to weak static disorder.  In the non-trivial region, the
edge state formation is even supported by disorder and, thus, the Fano
factor remains at the Poisson level.  In the trivial region, we witnessed a
slightly increased Fano factor.
\begin{figure}
\centerline{\includegraphics[width=\columnwidth]{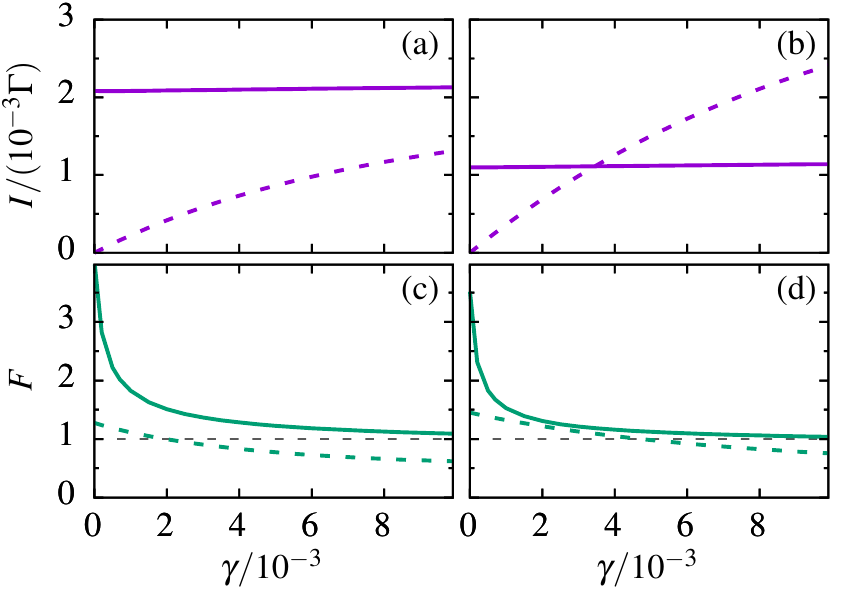}}
\caption{Influence of dissipation. (a,b) Time-averaged currents and (c,d)
corresponding Fano factors as a function of the dissipation rate
$\gamma$ for the amplitudes $qE/\omega=3.5$ and $qE/\omega=7$.  The left
and right column corresponds to the respective column of Fig.~\ref{fig:cdt},
i.e., (a,c) $\delta \tau=0.2 \tau_0$ and $b=0.4 a$, while (b) and (d)
correspond to the complementary case $\delta \tau=-0.2 \tau_0$ and $b=0.6
a$.  Solid lines mark topologically trivial cases, while dashed lines
correspond to non-trivial topology.  The driving frequency is
$\omega=5\tau_0$, while the chain-lead coupling reads $\Gamma=5\tau_0$.}
\label{fig:dissipation}
\end{figure}

Here we investigate the impact of a dynamic disorder stemming from the
interaction of each site with a respective heat bath via the population
operators $c_n^\dagger c_n$.  For weak coupling, we use a simple
description with a Lindblad operator \cite{Breuer2003a} with equal coupling
strengths and modify the Liouvillian according to
\begin{equation}
\mathcal{L} \to \mathcal{L} + \gamma\sum_n \mathcal{D}(c_n^\dagger c_n) ,
\end{equation}
where $\mathcal{D}$ is the Lindblad form defined after Eq.~\eref{eq:meq}.

Figures~\ref{fig:dissipation}(a) and \ref{fig:dissipation}(b) depicts how
the current changes upon increasing the dissipation strength for two
selected driving amplitudes.  We focus on the two complementary parameter
sets used in Fig.~\ref{fig:cdt} and select two particular driving
amplitudes, one corresponding to trivial topology (solid lines), the other
to non-trivial topology (dashed lines).  For trivial topology, the current
is rather insensitive to weak dissipation.  The main reason for this is
that in the trivial region, the transport occurs via the delocalized
eigenstates of the chain while coherences between these states play a minor
role \cite{Benito2016a}.  Accordingly, decoherence is not a relevant issue.
For non-trivial topology, by contrast, the current grows with an increasing
dissipation strength $\gamma$.  A physical picture for this behavior is the
direct transport between edge states.  Since the splitting of the edge
state doublet is exponentially small, the current is rather weak.  Then
dissipative transitions turn out to be rather beneficial for the electron
transport.

In contrast to the current, shot noise is affected by dissipation in the
same way as can be appreciated in Figs.~\ref{fig:dissipation}(c) and
\ref{fig:dissipation}(d).  For both trivial and non-trivial topology,
dissipation reduces the Fano factor which soon assumes value
close to the Poissonian $F=1$.  This means that measuring the topological
phase diagram via the Fano factor (see Fig.~\ref{fig:phasediagram}) will
require samples with very good coherence properties such that
$\gamma\lesssim10^{-3}\tau_0$, a value that seems feasible with present
quantum dot technology \cite{Forster2014a}.

\section{Conclusions}
\label{sec:conclusions}

We have investigated the influence of an AC driving on the current through
a SSH chain whose first and last site is coupled to an electron source and
drain, respectively.  Owing to their topological properties and the
corresponding presence of edge states, such chains have potential
applications in quantum information processing.  In the present case the
topological properties can be controlled in a very flexible manner via
driving frequency and amplitude.  In topologically non-trivial parameter
regions, edge states emerge and significantly influence the Fano factor of
the current.  In turn, the Fano factor may be used to measure the
topological phase diagram.

For the computation of the shot noise, we started from a generalized master
equation which leads to a numerical propagation scheme which, however, is
not very efficient for large system sizes.  To circumvent this problem, we
have developed a matrix-continued fraction method which is applicable
whenever the time-dependence of the current expectation value is weak, as
is the case for our model.

Within a high-frequency approximation, we have mapped the driven chain to
an effective time-independent model whose tunnel matrix elements are
dressed by Bessel functions.  Then in the topologically trivial region, the
transport occurs via many mutually exclusive channels. As is typical for
such mechanisms, we found super Poissonian shot noise.  In the non-trivial
region, by contrast, Poissonian long-distance tunneling between a pair of
edge states dominates.  At the zeros of Bessel functions, the effective
tunnel matrix elements and, thus, the current, vanish.  As an interesting
feature of driving-induced edge state blockade, not only the behavior of
the Fano factor, but also the shape of the current suppressions depends on
topology.

In summary, we have shown that AC fields not only allow one to tune the
topological properties of a SSH chain, but also that shot noise
measurements may serve for detecting Floquet topological transitions. Such
measurements may be an essential ingredient for testing and gauging setups
with applications in quantum information processing.

\ack
We would like to thank \'Alvaro G\'omez-Le\'on and Miguel Bello for helpful
discussions.
This work was supported by the Spanish Ministry of Economy and
Competitiveness via Grant No.\ MAT2014-58241-P and by the DFG via SFB~689.

\appendix
\section{AC components of the current}
\label{sec:I(t)}

\begin{figure}
\centerline{\includegraphics[width=\columnwidth]{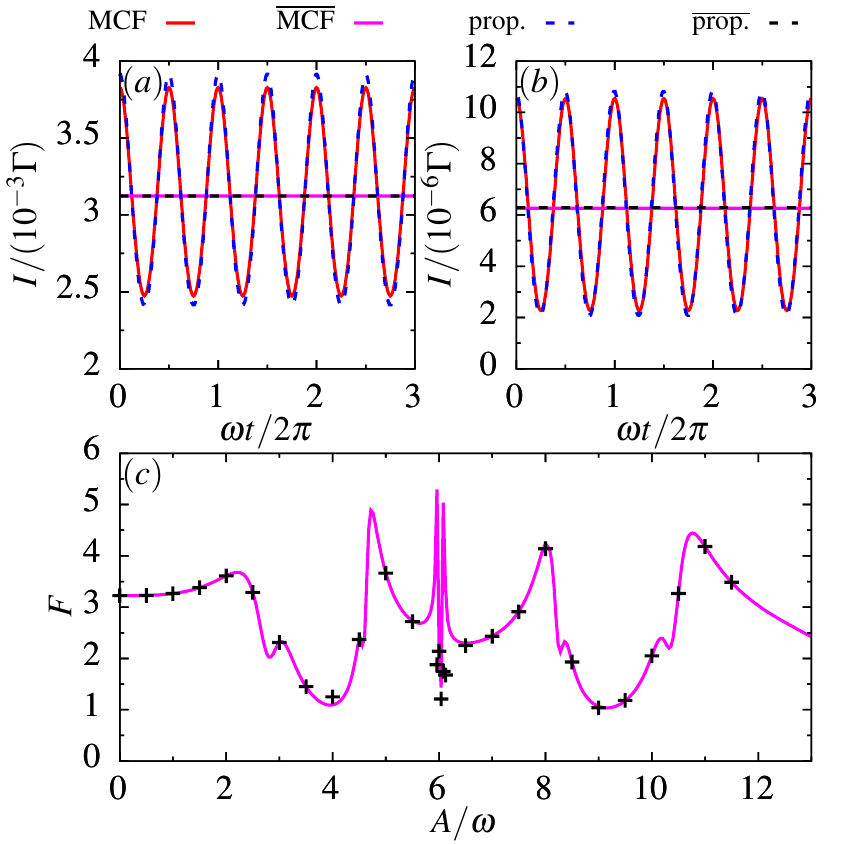}}
\caption{Time dependence of the current expectation value in the steady
state for a chain with length $N=14$, (a) $\delta \tau=0.2\tau_0$, $b=0.4
a$ and (b) $\delta \tau=-0.2\tau_0$, $b=0.6 a$ obtained by means of
propagation (prop.) and matrix-continued fractions (MCF). The respective
average values are marked by horizontal lines.  (c) Fano factor computed
with matrix-continued fractions (solid line) compared to the
result obtained by numerical propagation (dots) as a function of the
amplitude of the driving for $\delta \tau=-0.2\tau_0$ and $b=0.6 a$. The
driving frequency is $\omega=5\tau_0$, while the chain-lead coupling reads
$\Gamma=5\tau_0$.}
\label{fig:I(t)}
\end{figure}

The matrix-continued fraction method for the computation of the average
current significantly reduces the computational effort as compared to the
numerical propagation of the equations of motion \eref{eq:F0} and
\eref{eq:F1perp}, at least for large and intermediate driving frequencies
and for parameters that lead to current blockade.  To derive the former
method, however, we had to assume that in Eq.~\eref{eq:eqM}, the
time-dependent current $I(t)$ can be replaced by its time average Here we
test this assumption and show in Figs.~\ref{fig:I(t)}(a) and
\ref{fig:I(t)}(b) the time-dependence of $I(t)$ in the steady-state limit.
For typical driving parameters, we find that it possesses an appreciable ac
component even though it is always smaller than the time average.  In
Fig.~\ref{fig:I(t)}(c) we compare the results for the Fano factor computed
with matrix-continued fractions and by numerical propagation.  We find
that, despite the neglected time-dependence, both results agree rather
well.  This justifies the approximation made in Sec.~\ref{sec:mcf}.

\section*{References}
\providecommand{\newblock}{}

\end{document}